\documentclass[12pt,a4paper,oneside]{article}
\usepackage{a4wide}
\usepackage{amsmath}
\usepackage{graphicx}
\usepackage{rotating}
\usepackage{ifpdf}

\begin{document}

\textheight 21.0cm
\textwidth 16cm
\sloppy
\oddsidemargin 0.0cm \evensidemargin 0.0cm
\topmargin 0.0cm

%\begin{document}
\setlength{\parskip}{0.45cm}
\setlength{\baselineskip}{0.75cm}

%XXXXXXXXXXXXXXXXXXXXXXXXXXXXXXXXXXXXXX

%XXXXXXXXXXXXXXXXXXXXXXXXXXXXXXXXXXXXXX

\begin{titlepage}
\setlength{\parskip}{0.25cm}
\setlength{\baselineskip}{0.25cm}
\begin{flushright}
DO-TH 08/01\\
\vspace{0.2cm}
January 2008
\end{flushright}
\vspace{1.0cm}
\begin{center}
\Large
{\bf On the role of heavy flavor parton distributions}\\ 
{\bf  at high energy colliders}
\vspace{1.5cm}

\large
M.~Gl\"uck, P.~Jimenez-Delgado, E.\ Reya, C.~Schuck
\vspace{1.0cm}

\normalsize
{\it Universit\"{a}t Dortmund, Institut f\"{u}r Physik}\\
{\it D-44221 Dortmund, Germany} \\

\vspace{1.5cm}
\end{center}

\begin{abstract}
\noindent 
We compare `fixed flavor number scheme' (FFNS) and 
 `variable flavor number scheme' (VFNS) parton model predictions
at high energy colliders.  Based on our recent LO-- and NLO--FFNS
dynamical parton distributions, we generate radiatively two sets
of VFNS parton distributions where also the heavy quark flavors
$h=c,b,t$ are considered as massless partons within the nucleon.
By studying the role of these distributions in the production of
heavy particles ($h\bar{h},\, t\bar{b},\, hW^{\pm}$, Higgs--bosons,
etc.) at high energy $ep$, $p\bar{p}$ and $pp$ colliders, 
we show that the VFNS predictions are compatible with the FFNS ones
(to within about 10--20\% at LHC, depending on the process) when the
invariant mass of the produced system far exceeds the mass of the
participating heavy quark flavor.  
\end{abstract}
\end{titlepage}

%XXXXXXXXXXXXXXXXXXXXXXXXXXXXXXXXXXXXXXXXXXXXXXXXXXXXXXXXXXXXXXXXX
%MAIN PART
%XXXXXXXXXXXXXXXXXXXXXXXXXXXXXXXXXXXXXXXXXXXXXXXXXXXXXXXXXXXXXXXX
In a recent publication \cite{ref1} we updated the dynamical leading
order (LO) and next--to--leading order (NLO) parton 
distributions of \cite{ref2}.  These analyses were undertaken within
the framework of the so called  `fixed flavor number scheme' (FFNS)
where, besides the gluon, only the light quarks $q=u,d,s$ are considered
as genuine, i.e.\ massless partons within the nucleon.  
This factorization scheme is fully predictive in the heavy quark
$h=c,b,t$ sector where the heavy quark flavors are produced entirely
perturbatively from the initial light quarks and gluons -- as required
experimentally, in particular in the threshold region.  However, even
for very large values of $Q^2,\, Q^2\gg m_{c,b}^2$, these FFNS
predictions are in remarkable agreement with DIS data \cite{ref1,ref2} 
and, moreover, are perturbatively stable, despite the common belief that
`non--collinear' logarithms $\ln(Q^2/m_h^2)$ have to be resummed.

In many situations, however, calculations within this factorization
scheme become unduly complicated.  For example, the single top production
process at hadron colliders via $W$--gluon fusion requires the calculation
of the subprocess $ug\to d\,t\,\bar{b}$ at LO and of 
$ug\to d\,t\,\bar{b}\,g$ 
etc.\ at NLO\@.  It thus becomes expedient to consider for such calculations
the so called  `variable flavor number scheme' (VFNS) where also the
heavy quarks $h=c,b,t$ are taken to be massless partons within the 
nucleon.  In this scheme, the above FFNS calculations simplify 
considerably, i.e.\ one needs merely $ub\to dt$ at LO and $ub\to dtg$
etc.~at the NLO of perturbation theory \cite{ref3}.  The VFNS is
characterized by increasing the number $n_f$ of massless partons by one
unit at $Q^2=m_h^2$ starting from $n_f=3$ at $Q^2=m_c^2$, i.e.\ 
$c(x,m_c^2) = \bar{c}(x,m_c^2)=0$.  The matching conditions at LO and 
NLO are fixed by \mbox{continuity} relations \cite{ref4} at the respective
thresholds $Q^2=m_h^2$.  Thus the  `heavy' $n_f>3$ quark \mbox{distributions}
are perturbatively uniquely generated from the $n_f\! -\!1$ ones via the 
massless renormalization group $Q^2$--evolutions.
%resubmission%%%%%%%%%%%%%%%%%%%%%%%%%%%%%%%%%%%%%%%%%%%%%%%%%%%%%
The running strong coupling can be approximated by the common NLO
   `asymptotic' solution
%%%%%%%%%%%%%%%%%%%%%%%%%%%%Eq.(1)%%%%%%%%%%%%%%%%%%%%%%%%%%%%%%%%%%
\begin{equation}
\frac{\alpha_s(Q^2)}{4\pi} 
  \simeq \frac{1}{\beta_0\ln (Q^2/\Lambda^2)} 
    - \frac{\beta_1}{\beta_0^3} \,\,
        \frac{\ln\ln (Q^2 / \Lambda^2)}{[\ln (Q^2/\Lambda^2)]^2}
\end{equation}
%%%%%%%%%%%%%%%%%%%%%%%%%%%%%%%%%%%%%%%%%%%%%%%%%%%%%%%%%%%%%%%%%%%%
with $\beta_0=11-2n_f/3$ and $\beta_1=102-38n_f/3$, which turns out to
be sufficiently accurate \cite{ref1} for our relevant $Q^2$-region,
$Q^2$ \raisebox{-0.1cm}{$\stackrel{>}{\sim}$} 2 GeV$^2$.
Since $\beta_{0,1}$ are not continuous for different flavor numbers
$n_f$, the continuity of $\alpha_s(Q^2)$ requires to choose different
values for the integration constant $\Lambda$ for different $n_f$,
$\Lambda^{(n_f)}$, which are fixed by the common $\alpha_s(Q^2)$
matchings at the flavor thresholds $Q=m_{c,b,t}$. Choosing $m_c=1.3$ 
GeV, $m_b=4.2$ GeV and $m_t=175$ GeV, one obtains 
$\Lambda_{\overline{\rm MS}}^{(4,5,6)}= 269.7$, 184.5, 72.9 MeV
according to our dynamical NLO($\overline{\rm MS}$) fit \cite{ref1}
which resulted in $\alpha_s(M_Z^2)= 0.1145$. In LO, where 
$\beta_1\equiv 0$, we obtained \cite{ref1} $\Lambda_{LO}^{(4,5,6)}=
181.8$, 138.3, 70.1 MeV corresponding to $\alpha_s(M_Z^2)=0.1263$. 
%%%%%%%%%%%%%%%%%%%%%%%%%%%%%%%%%%%%%%%%%%%%%%%%%%%%%%%%%%%%%%%%%%%%

Our choice for the input of the  `heavy' VFNS distributions are the LO
and NLO dynamical FFNS distributions \cite{ref1} at $Q^2=m_c^2$.  The
VFNS predictions at scales $Q^2\gg m_h^2$ should become insensitive
to this, somewhat arbitrary, input selection \cite{ref5}  whose virtues
are the fulfillment of the standard sum--rule constraints together with
reasonable shapes and sizes of the various input distributions. 
%%%%%%resubmission%%%%%%%%%%%%%%%%%%%%%%%%%%%%%%%%%%%%%%%%%%%%%%%%%%%
As we shall see, this expectation is based on the fact that at 
$Q^2\gg m_h^2$ the VFNS distributions are dominated by their radiative
evolution rather than by the specific input at $Q^2=m_h^2$.  In other
words, because of the long evolution distance, input differences get
`evolved away' at $Q^2\gg m_h^2$ where the universal perturbative
QCD splittings dominate.
%%%%%%%%%%%%%%%%%%%%%%%%%%%%%%%%%%%%%%%%%%%%%%%%%%%%%%%%%%%%%%%%%%%% 

As a first test of the VFNS  `heavy' quark distributions we consider 
charm and bottom electroproduction processes, since deep inelastic 
structure functions play an instrumental role in determining parton
distributions.  In Figs.~1 and 2 we compare the VFNS with the FFNS 
predictions for $F_2^c(x,Q^2)$ and $F_2^b(x,Q^2)$, respectively, 
using\footnote[1]{Notice that here and in the following 
$\mu_R=\mu_F\equiv\mu$
where $\mu_R$ and $\mu_F$ are the renormalization and factorization
scales, respectively.  This choice is dictated by the fact that our 
(and all other presently available) parton distributions were determined
and evolved with $\mu_R=\mu_F$, i.e.\ with the commonly adopted standard
evolution equations.} 
$\mu^2=Q^2+4m_{c,b}^2$ for the FFNS although our results are not very
sensitive to this specific choice of the factorization and renormalization
scales. As usual, $\mu^2=Q^2$ in the VFNS\@.  Notice furthermore that the 
NLO--VFNS predictions for $xh$ (short--dashed curves) are very similar
to the ones for $(2e_h^2)^{-1}F_2^h$ (dashed curves) despite the fact
that $(2e_h^2)^{-1}F_2^h=(1+\alpha_sC_q)
\otimes h+\frac{1}{2}\alpha_s C_g\otimes g$,
i.e.\ the ${\cal{O}}(\alpha_s)$ quark and gluon contributions almost
cancel.  As expected \cite{ref6} the discrepancies between the predictions
for $xh(x,Q^2)$ in the VFNS and for $(2e_h^2)^{-1}F_2^h(x,Q^2)$ in the 
FFNS in the relevant kinematic region (small $x$, large $Q^2$) never
disappear and can amount to as much as about 30\% at very small--$x$, 
even at $W^2\equiv Q^2(\frac{1}{x}-1)$ far above threshold, i.e.\ 
$W^2\gg W_{th}^2=(2m_h)^2$.  This is due to the fact that here the 
ratio of the threshold energy $W_{th}\equiv\sqrt{\hat{s}_{th}}$ of the
massive subprocess ($\gamma^*g\to h\bar{h}$, etc.) and the mass of the 
produced heavy quark $\sqrt{\hat{s}_{th}}/m_h=2$ is not sufficiently high
to exclude significant contributions from the threshold region.  Even
for the lightest heavy quark, $h=c$, such non--relativistic 
$(\beta_c=|\vec{p}_c|/E_c$ \raisebox{-0.1cm}{$\stackrel{<}{\sim}$} 0.9)
threshold region contributions to $F_2^h(x,Q^2)$ are sizeable for 
$W^2$ \raisebox{-0.1cm}{$\stackrel{<}{\sim}$} $10^6$ GeV$^2$ due to
significant $\beta_c<0.9$ contributions, and the situation becomes worse
for $h=b$ (cf.\ Fig.~4 of \cite{ref6}).  This is in contrast to processes
where one of the produced particles is much heavier than the other one,
like the weak $CC$ contribution \cite{ref7,ref8} $W^+g\to t\bar{b}$ to
$F_2^{CC}$.  Here $\sqrt{\hat{s}_{th}}/m_b=(m_t+m_b)/m_b\gg 1$ and the
extension of the threshold region where $\beta_{\bar{b}}$ 
 \raisebox{-0.1cm}{$\stackrel{<}{\sim}$} 0.9, being proportional to 
$m_b/\sqrt{\hat{s}_{th}}\ll 1$, is strongly reduced as compared to
$m_h/(2m_h)=0.5$ in the former case of $h\bar{h}$ production.  Thus the
single top production rate in $W^+g\to t\bar{b}$ is dominated by the 
(beyond--threshold) relativistic region where $\beta_{\bar{b}}>0.9$ and 
therefore is expected to be well approximated by $W^+b\to t$ where $b$ is
an effective massless parton within the nucleon.  In Fig.~3 we compare
the LO FFNS \cite{ref7,ref8} predictions for 
$\frac{1}{2}F_{2,t\bar{b}}^{CC}(x,Q^2)$ with the corresponding VFNS ones
for $\xi b(\xi,Q^2+m_t^2)$ where the latter refers to the $W^+b\to t$
transition using the slow rescaling variable \cite{ref9} 
$\xi=x(1+m_t^2/Q^2)$ with $m_t=175$ GeV\@.  For $F_{2,t\bar{b}}^{CC}(x,Q^2)$
we used $\mu_R^2=\mu_F^2\equiv\mu^2=Q^2+(m_t+m_b)^2$. (Notice that
the fully massive NLO FFNS QCD corrections to $W^+g\to t\bar{b}$ are 
unfortunately not available in the literature.)  As expected the
differences between the two schemes are here less pronounced than in
the case of $c\bar{c}$ and $b\bar{b}$ electroproduction in Figs.~1 
and 2. These results indicate that one may resort to the simpler VFNS
with its massless $h(x,\mu^2)$ distributions to estimate rather reliably
the production rates of heavy quarks, gauge bosons, Higgs scalars, etc.
at Tevatron and LHC energies.

As a next test of these VFNS distributions we therefore turn to hadronic
$W^{\pm}$ production and present in Fig.~4 their NLO predictions for
$\sigma(p\bar{p}\to W^{\pm}X)$ as compared to the data 
\cite{ref10,ref11,ref12,ref13,ref14} and to predictions based on the 
NLO CTEQ6.5 distributions \cite{ref15}.  Also shown in this figure is
a comparison of our LO FFNS and VFNS predictions. Although quantitatively
slightly different, the dominant light quark contributions in the FFNS
($u\bar{d}\to W^+,\, u\bar{s}\to W^+$, etc.) are due to the same
subprocesses as in the VFNS, but the relevant heavy quark contributions
have been calculated via $g\bar{s}(\bar{d})\to\bar{c}W^+$, $gu\to bW^+$, 
etc.\ as compared to $c\bar{s}(\bar{d})\to W^+$, $\bar{b}u\to W^+$ etc.\ in the
VFNS\@. Here we again expect the VFNS with its effective massless `heavy'
quark distributions $h(x,\mu^2)$ to be adequate, since non--relativistic
contributions from the threshold region in the FFNS are suppressed due
to $\sqrt{\hat{s}_{th}}/m_{c,b}\simeq M_W/m_{c,b}\gg 1$. The LO gluon
induced heavy quark contributions to $W^{\pm}$ production in the FFNS
are obtained from a straightforward calculation of the differential cross
section \cite{ref16} $d\hat{\sigma}(\hat{s})/d\hat{t}$ which yields
%%%%%%%%%%%%%%%%%%%%%%%%%%%%%%%%%%Eq.2%%%%%%%%%%%%%%%%%%%%%%%%%%%%%%%%%%
\begin{eqnarray}
\hat{\sigma}(\hat{s})^{gs\to cW^-}  = \frac{G_F}{\sqrt{2}} \,\, 
   \frac{\alpha_s(\mu^2)}{6}\,\, |V_{cs}|^2 \,\,\frac{M_W^2}{\hat{s}}\,\!\!\!\!\!
    & & \!\!\!\!\Bigg\{ \left( 1+\frac{m_c^2}{2M_W^2}\right)
     \Big[ \frac{\sqrt{\lambda}}{2}\, (1+7\Delta)
\\
& & +(1-2\Delta+2\Delta^2)
       \ln \frac{1-\Delta+\sqrt{\lambda}}{1-\Delta -\sqrt{\lambda}}\Big]
         -\frac{m_c^2}{M_W^2}\sqrt{\lambda} \Bigg\}\nonumber
\end{eqnarray}
%%%%%%%%%%%%%%%%%%%%%%%%%%%%%%%%%%%%%%%%%%%%%%%%%%%%%%%%%%%%%%%%%%%%%%
where
\begin{displaymath}
 \Delta = \frac{M_W^2-m_c^2}{\hat{s}}\quad, \quad\quad
 \lambda = \Big[ 1-\frac{(m_c+M_W)^2}{\hat{s}}\Big]\,\,
           \Big[ 1-\frac{(m_c-M_W)^2}{\hat{s}}\Big]\,\, ,
\end{displaymath}
$\alpha_s(\mu^2)=4\pi/[9\ln(\mu^2/\Lambda_{\rm LO}^{(3)})]$ and the
relevant CKM matrix element(s) $V_{qq'}$ are taken from \cite{ref17}.
The corresponding total $W^{\pm}$ hadronic production cross section
relevant for Fig.~4 is then given by
%%%%%%%%%%%%%%%%%%%%%%%%%%%%%%%%%Eq.3%%%%%%%%%%%%%%%%%%%%%%%%%%%%%%%
\begin{equation}
\sigma^{p\bar{p}\to cW^{\pm}X}(s) =
  \int_{\tau}^1 dx_1 \int_{\tau/x_1}^1 dx_2\, 
    [g(x_1,\mu^2)s(x_2,\mu^2)+(1\leftrightarrow2)]\, 
      \hat{\sigma}(x_1 x_2 s)
\end{equation}
%%%%%%%%%%%%%%%%%%%%%%%%%%%%%%%%%%%%%%%%%%%%%%%%%%%%%%%%%%%%%%%%%%%%%
where $s(x,\mu^2)=\bar{s}(x,\mu^2)$ with $\mu^2={\cal{O}}(M_W^2)$
and $\tau=(m_c+M_W)^2/s$.  Unfortunately, the NLO(${\overline{\rm MS}}$)
corrections to this (massive quark) FFNS cross section are again not
available in the literature.  Only quantitative LO and NLO results for
the analogous process $gb\to tW^-$ have been presented in \cite{ref18},
but questioned in \cite{ref19}.  Here we just mention that we fully
confirm the LO results for $Wt$ production obtained in \cite{ref19}
at Tevatron and LHC energies.  Taking into account that the 
$K \equiv$ NLO/LO factor is expected \cite{ref19} to be in the range
of 1.2 -- 1.3, our LO--FFNS predictions in Fig.~4 imply equally
agreeable NLO predictions as the (massless quark) NLO--VFNS ones
\cite{ref20} shown by the solid and dashed--dotted curves in Fig.~4.

In Table 1 we present our VFNS and FFNS predictions for $W^{\pm}$
production at LHC and compare the relevant subprocess contributions
to $\sigma(pp\to W^{\pm} X)$ at $\sqrt{s}=14$ TeV. The light quark
fusion contributions in the $ud$ and $us$ sector turn out to be
somewhat larger in the FFNS than in the VFNS which is due to the fact
that the $u,d,s$ (and the gluon) distributions are evolved for fixed
$n_f=3$ in the FFNS\@.  More interesting, however, are the subprocesses
involving heavy quarks.  Here the LO--VFNS predictions are compatible,
to within less than 15\%, with the LO--FFNS predictions based on the gluon
induced fixed order in $\alpha_s$ subprocesses $gu\to bW$, 
$gd\to cW$ and in particular on the sizeable CKM non--suppressed
$gs\to cW$ contribution.  As mentioned above, the NLO corrections
to these latter heavy quark contributions cannot be calculated for
the time being.  However, since these contributions amount to about
only 15\% of the total FFNS results for $W^{\pm}$ production (being
dominated by the light $ud$ and $us$ fusions in Table 1), we can
{\em safely} employ the expected \cite{ref19} $K$ factor of 
$K\simeq 1.2$ for the relevant $gd\to cW$ and $gs\to cW$ 
LO contributions in Table 1 for obtaining the total NLO--FFNS
predictions without committing any significant error.  The resulting
total rate for $W^++W^-$ production at LHC of $192.7 \pm 4.7\,\, {\rm nb}$
is comparable to our NLO--VFNS prediction in Table 1 of 
$186.5\pm 4.9$ nb where we have added the $\pm 1\sigma$ uncertainties 
implied by our dynamical parton distributions 
\cite{ref1}.\footnote[2]{Using  `standard' FFNS parton distributions 
\cite{ref1} instead of the dynamical ones for generating the VFNS 
distributions, the dynamical NLO--VFNS prediction of 186.5 nb slightly
increases to 190.7 nb.}
This latter prediction reduces to 181.0 nb when using the smaller scale
$\mu^2 =M_W^2/4$.  The scale uncertainties of our predictions are
defined by taking $M_W/2\leq \mu\leq 2\,M_W$, using $M_W=80.4$ GeV,
which gives rise to the upper limits ($\mu=2M_W$) and lower limits
$(\mu=M_W/2)$ of our predicted cross sections.  In this way we obtain
the following total uncertainty estimates of our NLO predictions at
LHC:
%%%%%%%%%%%%%%%%%%%%%%%%%%%%%%%%%%Eq.4%%%%%%%%%%%%%%%%%%%%%%%%%%%%%%
\begin{equation}
\sigma(pp\to W^+ +W^-+X) = \Bigg\{
\renewcommand{\arraystretch}{1.5}
 \begin{array}{c} 
  186.5 \pm 4.9_{\rm pdf}\,\,_{-5.5}^{+4.8}
            \mid_{\rm scale}\quad {\rm nb}\,\, , \quad {\rm VFNS} \\
  192.7 \pm 4.7_{\rm pdf}\,\,_{-4.8}^{+3.8}
            \mid_{\rm scale}\quad {\rm nb}\,\, , \quad {\rm FFNS} \end{array}
\end{equation}
%%%%%%%%%%%%%%%%%%%%%%%%%%%%%%%%%%%%%%%%%%%%%%%%%%%%%%%%%%%%%%%%%%%%
and, for completeness, at LO
%%%%%%%%%%%%%%%%%%%%%%%%%%%%%%%%%Eq.5%%%%%%%%%%%%%%%%%%%%%%%%%%%%%%%
\begin{equation}
\sigma(pp\to W^+ +W^-+X) = \Bigg\{
\renewcommand{\arraystretch}{1.5}
 \begin{array}{c} 
  162.1 \pm 3.9_{\rm pdf}\,\,_{-21.8}^{+20.3}
            \mid_{\rm scale}\quad {\rm nb}\,\, , \quad {\rm VFNS} \\
  166.7 \pm 4.0_{\rm pdf}\,\,_{-19.0}^{+17.3}
            \mid_{\rm scale}\quad {\rm nb}\,\, , \quad {\rm FFNS} \end{array}
\end{equation}
%%%%%%%%%%%%%%%%%%%%%%%%%%%%%%%%%%%%%%%%%%%%%%%%%%%%%%%%%%%%%%%%%%%%%%%%%
where the subscript pdf refers to the $1\sigma$ uncertainties of our 
parton distribution functions \cite{ref1}.  For comparison, the NLO--VFNS
prediction of CTEQ6.5 \cite{ref15} is 202 nb with an uncertainty of 8\%,
taking into account a pdf uncertainty of slightly more than $2\sigma$.
Similarly, MRST \cite{ref21} predict about 194 nb. From these results
we conclude that, for the time being, the total $W^{\pm}$ production
rate at LHC can be safely predicted within an uncertainty of about 10\%
irrespective of the factorization scheme.

It is also interesting to study the dependence of the FFNS predictions
for the contributions to $W^{\pm}$ production involving heavy quarks on
the chosen scale $\mu$ as shown in Figs.~5 and 6.  In these figures we
compare the $gs \to cW$ initiated production rates in the FFNS with the
quark fusion $cs\to W$ ones in the VFNS and similarly the $gd\to cW$
ones with the $cd \to W$ fusion, respectively.  These factorization 
scheme dependencies are rather mild for the LO--FFNS predictions, in contrast
to the situation for the LO--VFNS predictions which stabilize, as expected,
at NLO.  The mild $\mu$ dependence is similar to the situation encountered
in $tW$ production \cite{ref19} via the subprocess $gb\to tW^-$.

A similar situation where the invariant mass of the produced system
sizeably exceeds the mass of the participating heavy quarks is encountered
in (heavy) Higgs $H$ production accompanied by two heavy $b$--quarks, for
example.  Here $H=H_{\rm SM}^0;\,\, h^0,\, H^0,\, A^0$ denotes the 
SM Higgs--boson or a light scalar $h^0$, a heavy scalar $H^0$ and a 
pseudoscalar $A^0$ of supersymmetric theories with $M_H$
\raisebox{-0.1cm}{$\stackrel{>}{\sim}$} 100 GeV.  In the FFNS the
dominant production mechanism starts with the LO subprocess 
$gg\to b\bar{b}H$ $(q\bar{q}\to b\bar{b}H$ is marginal), to be compared with
the $b\bar{b}$ fusion subprocess in the VFNS starting with $b\bar{b}\to H$
at LO\@.  Again, $\sqrt{\hat{s}_{th}}/m_b=(2m_b+M_H)/m_b\simeq M_H/m_b\gg 1$
in the FFNS which indicates that the simpler LO and NLO(NNLO) VFNS 
$b\bar{b}$ fusion subprocesses do provide reliable predictions.
Within the scale uncertainties it turns out that the 
FFNS and VFNS predictions at NLO are compatible 
\cite{ref22,ref23,ref24,ref25},
%%%%%%%%resubmission%%%%%%%%%%%%%%%%%%%%%%%%%%%%%%%%%%%%%%%%%%%%%%%%%
using the MRST2002 and CTEQ6 parametrizations of the relevant parton
distributions \cite{ref21,ref26}.
%%%%%%%%%%%%%%%%%%%%%%%%%%%%%%%%%%%%%%%%%%%%%%%%%%%%%%%%%%%%%%%%%%%%%
 This result holds for scale choices
$\mu_{R,F}=(\frac{1}{8}\,{\rm to}\,\frac{1}{2})\sqrt{\hat{s}_{th}}$ with
$\sqrt{\hat{s}_{th}}/4$ being considered as a suitable `central' choice
in $b(x,\mu_F^2)$ for calculations based on the $b\bar{b}$ fusion 
process in the VFNS\@.\footnote[3]{The
{\em independent} variation of $\mu_F$ and $\mu_R$ considered in
\cite{ref22,ref23,ref24,ref25} is, as mentioned before, not strictly
compatible with the utilized parton distributions determined and evolved
according to $\mu_R=\mu_F$.}
It should, however, be mentioned that the VFNS rates somewhat exceed 
\cite{ref22,ref23,ref24,ref25} the corresponding FFNS Higgs--boson
production rates by about$^3$ 10--20\%.

Finally let us note that all our results and comparisons 
concerning the VFNS hold irrespective of the specific parametrizations
used for the `heavy' $h(x,\mu^2)$ distributions: when comparing our 
VFNS distributions, generated from using our dynamical distributions
\cite{ref1} as input, with the ones of 
%%%%%%resubmission%%%%%%%%%%%%%%%%%%%%%%%%%%%%%%%%%%%%%%%%%%%%%%%%
CTEQ6 \cite{ref26} or
%%%%%%%%%%%%%%%%%%%%%%%%%%%%%%%%%%%%%%%%%%%%%%%%%%%%%%%%%%%%%%%%%%
CTEQ6.5 \cite{ref15} the relevant
ratios $c(x,M_W^2)_{\rm CTEQ}/c(x,M_W^2)_{\rm GJR-VFNS}$ and 
$b(x,m_t^2)_{\rm CTEQ}/b(x,m_t^2)_{\rm GJR-VFNS}$ vary, for $10^{-4}$
\raisebox{-0.1cm}{$\stackrel{<}{\sim}$} x 
\raisebox{-0.1cm}{$\stackrel{<}{\sim}$} 0.1, at most between 
0.9 -- 1.1 at LO and NLO. 
%%%%%%resubmission%%%%%%%%%%%%%%%%%%%%%%%%%%%%%%%%%%%%%%%%%%%%%%%%
Similar results hold when using other VFNS distributions, e.g., those
of \cite{ref5}.  This is illustrated more quantitatively in Fig.~7
where we compare our $c$-- and $b$--distributions, together with the
important gluon--distribution, with the ones of CTEQ6 \cite{ref26}
and CTEQ6.5 \cite{ref15} in the sea-- and gluon--relevant $x$--region
$x$ \raisebox{-0.1cm}{$\stackrel{<}{\sim}$} $0.3$ at $Q^2=M_W^2$.
The ratios for the light $u$-- and $d$--distributions are even closer
to 1 than the ones shown in Fig.~7, typically between 0.95 and 1.05
which holds in particular for the CTEQ6 distributions when compared
to our ones. Incidentally the VFNS under consideration and commonly
used \cite{ref5,ref26} is also referred to as the zero--mass VFNS.
Sometimes one uses an improvement on this, now known as the 
general--mass VFNS 
%%%%%%%2.resubmission%%%%%%%%%%%%%%%%%%%%%%%%%%%%%%%%%%%%%%%%%%%%%%%%%
\cite{ref15,ref21,ref27,ref28,ref29,ref30}, 
%%%%%%%%%%%%%%%%%%%%%%%%%%%%%%%%%%%%%%%%%%%%%%%%%%%%%%%%%%%%%%%%%%%%%%
where mass--dependent
corrections are maintained in the hard cross sections.
This latter factorization scheme interpolates between the strict
zero--mass VFNS, used in our evolution to $Q^2\gg m_h^2$, and the 
(experimentally required) FFNS used for our input at $Q^2=m_h^2$.
As expected and shown in Fig.~7, scheme (input) differences at lower
$Q^2={\cal{O}}(m_h^2)$ only marginally affect the asymptotic results
at $Q^2=M_W^2\gg m_{c,b}^2$ where the CTEQ6.5 parametrizations 
\cite{ref15} (corresponding to a general--mass VFNS) become very
similar to the ones of CTEQ6 \cite{ref26} and our GJR--VFNS (corresponding
to the zero--mass VFNS).
%%%%%%2.resubmission%%%%%%%%%%%%%%%%%%%%%%%%%%%%%%%%%%%%%%%%%%%%%%%%%%%
As stated repeatedly before, this is essentially due to the dominance 
of the large evolution effects over the minor differences involved at the
lower scales, e.g.\ at $Q^2={\cal{O}}(m_h^2)$.  These asymptotic similarities
are particularly relevant for the simplified (vanishing $m_{c,b}$) 
calculations of the production rates of very massive particles where
massive $c$-- and $b$--quark threshold region contributions are strongly
suppressed.  
%%%%%%%%%%%%%%%%%%%%%%%%%%%%%%%%%%%%%%%%%%%%%%%%%%%%%%%%%%%%%%%%%%%%

To summarize, we generated radiatively two sets of VFNS parton distributions,
based on 
our recent LO and NLO dynamical parton distributions \cite{ref1} obtained
in the FFNS. Within the VFNS additional heavy quark distributions $h(x,Q^2)
=\bar{h}(x,Q^2)$ are generated perturbatively via the common massless 
$Q^2$--evolution equations by imposing the boundary conditions 
$h(x,Q^2=m_h^2)=0$ for $h=c,b,t$.  We have confronted the VFNS and FFNS
predictions in situations where the invariant mass of the produced system
($h\bar{h},\, t\bar{b},\, cW,\, tW$, Higgs--bosons, etc.) does not exceed
or exceeds by far the mass of the participating heavy flavor.  In the 
former case (e.g.\ $F_2^c$ in deep inelastic $c\bar{c}$ production where
$\sqrt{\hat{s}_{th}}/m_c=2$) the VFNS predictions deviate from the FFNS
ones by up to about 30\% even at $Q^2\gg m_c^2$.  In the latter case 
(e.g.\ $F_{2,t\bar{b}}^{CC}$ in deep inelastic weak charged current 
$t\bar{b}$ production where $\sqrt{\hat{s}_{th}}/m_b\simeq m_t/m_b\gg 1$
these deviations are appreciably less, within about 10\%, which is within
the margins of renormalization and factorization scale uncertainties.
As a further example of the agreement between the VFNS and FFNS predictions
in situations where the invariant mass of the produced system far exceeds
$m_{c,b}$ we considered the corresponding $W^{\pm}$ boson production rates at
the Tevatron and at the large hadron collider (where e.g.\ 
$\sqrt{\hat{s}_{th}}/m_{c,b}\simeq M_W/m_{c,b}\gg 1$ for $cW$ and $bW$
production, respectively).  For $\sqrt{s}=14$ TeV the NLO--FFNS predicts 
$\sigma(pp\to W^+ +W^-+X)\simeq 192.7$ nb with an uncertainty of 5\%,
to be compared with the NLO--VFNS prediction of 186.5 nb and an uncertainty
of 6\%.  The cited uncertainties include also the scale uncertainties due
to varying the renormalization and factorization scales $\mu_R=\mu_F$
between $M_W/2$ and $2M_W$. (It should be emphasized again that the scale choice
$\mu_R=\mu_F$ is dictated by all presently available parton distributions
which have been determined and evolved according to $\mu_R=\mu_F$.) 
Furthermore, a similar agreement is obtained for hadronic (heavy)
Higgs--boson production when the dominant FFNS subprocess $gg\to b\bar{b}H$
(where $\sqrt{\hat{s}_{th}}/m_b=(2m_b+M_H)/m_b\gg 1)$ is compared with
the VFNS $b$--quark fusion subprocess ($b\bar{b}\to H$, etc.).

These results indicate that the simpler VFNS with its effective treatment
of heavy quarks ($c,b,t$) as massless partons can be employed for 
calculating processes where the invariant mass of the produced system is
sizeably larger than the mass of the participating heavy quark flavor.
The uncertainty of such calculations is process (and somewhat energy)
dependent when compared with the predictions of the FFNS where the effects
of finite heavy quark masses are nowhere neglected.  Taking into account
the uncertainties of parton distributions and scale choices as well, the
total $W^{\pm}$ production rate at LHC can be predicted within an
uncertainty of about 10\%, which becomes significantly smaller at the Tevatron.
Similarly the Higgs production rates at LHC are predicted with an 
uncertainty of 10--20\% where the VFNS production rates exceed the FFNS
ones by about 20\% at LHC.

A FORTRAN code (grid) containing our new LO and NLO($\overline{\rm MS}$)
light ($u,d,s;g$) and heavy ($c,b,t$) parton distributions in the 
VFNS, generated from our recent dynamical ones in the FFNS \cite{ref1},
can be obtained on request or directly from 
{\tt{http://doom.physik.uni-dortmund.de}}$\,$ .
\vspace{1.0cm}

\noindent{\large{\bf Acknowledgements}}\\
We thank J.~Campbell for a clarifying correspondence.  This work has been
supported in part by the  `Bundesministerium f\"ur Bildung und Forschung',
Berlin/Bonn.

%%%%%%%%%%%%%%%%%%%%%%%%%%%%%%Table 1%%%%%%%%%%%%%%%%%%%%%%%%%%%%%%%%%%%%%
%\newpage
\pagestyle{empty}
\begin{sidewaystable}[th]
\begin{center}
\renewcommand{\arraystretch}{1.5}
\begin{tabular}{|c|c|c|c||c|c|c|c|}
\multicolumn{8}{|c|}{$\sigma^{\textrm{pp}\rightarrow\textrm{W}X}$ (nb),\textrm{ } $\sqrt{\textrm{s}}$=14 TeV}\\\hline
\multicolumn{4}{|c||}{VFNS: NLO (LO)}&\multicolumn{4}{|c|}{FFNS: NLO (LO)}\\
               & $\textrm{W}^+$ & $\textrm{W}^-$ & $\textrm{W}^+ + \textrm{W}^-$ &
$\textrm{W}^+$ & $\textrm{W}^-$ & $\textrm{W}^+ + \textrm{W}^-$                  & \\\hline
  ud  &  87.7 (77.6) & 60.6 (52.6)  & 148.3 (130.3) &  93.3 (81.9) & 64.6 (55.7) & 157.9 (137.6) &     ud            \\
  us  &    3.9 (3.3) &  1.3 (1.2)  &   5.3 (4.5)   &  4.2 (3.5) & 1.5 (1.3) & 5.7 (4.8) &     us            \\
  ub  &    7.3 (7.0)$\times10^{-4}$ &  2.3 (2.2)$\times10^{-4}$  &   9.6 (9.2)$\times10^{-4}$   &
 \mbox{ -~ } (6.5)$\times10^{-4}$ & \mbox{ -~ } (1.8)$\times10^{-4}$ & \mbox{ -~ } (8.3)$\times10^{-4}$ & gu$\rightarrow$bW \\
  cd  &    1.3 (1.1) &  2.3 (2.0)  &   3.6 (3.1)   & \mbox{ -~~ } (1.0) & \mbox{ -~~ } (2.0) & \mbox{ -~~ } (3.0) & gd$\rightarrow$cW \\
  cs  &  14.7 (12.2) & 14.7 (12.2) &  29.4 (24.3)  & \mbox{ -~~  } (10.6) & \mbox{ -~~ } (10.6) & \mbox{ -~~ } (21.3) & gs$\rightarrow$cW \\
  cb  &   1.5 (1.4)$\times10^{-2}$ &  1.5 (1.4)$\times10^{-2}$  &   2.9 (2.7)$\times10^{-2}$   & - & - & - & - \\\hline
total & 107.5 (94.2) & 79.1 (67.9) & 186.5 (162.1) & $\simeq$ 111.4 (97.0) & $\simeq$ 81.2 (69.6) & $\simeq$ 192.7 (166.7) & total        \\\hline
\end{tabular}
\caption{NLO(LO) VFNS and FFNS predictions for $W^{\pm}$ production at LHC\@.
The FFNS parton distributions are taken from \cite{ref1} which form the basis
for generating the ones in the VFNS including the  `heavy' quark distributions
$c(x,\mu^2)=\bar{c}(x,\mu^2)$ and $b(x,\mu^2)=\bar{b}(x,\mu^2)$ taking $\mu=M_W$.  
The uncertainties implied by different scale choices are summarized in 
eqs.~(4) and (5).  The total NLO--FFNS rates have been obtained by adopting
an expected \cite{ref19} $K$--factor of 1.2 for the subleading gluon initiated
LO rates involving the heavy $c$ and $b$ quarks.}
\end{center}
\end{sidewaystable}
%%%%%%%%%%%%%%%%%%%%%%%%%%%%%%%%%%%%%%%%%%%%%%%%%%%%%%%%%%%%%%%%%%%%%%%%%5
\clearpage
%%%%%%%%%%%%%%%%%%%%%%%%%%%%%%%%%%%%%%%%%%%%%%%%%%%%%%%%%%%%%%%%%%%%%%%%%%
%%%%%%%%%%%%%%%%%%%%%%%%%%%%%%%%%%%%%%%%%%%%%%%%%%%%%%%%%%%%%%%%%%%%%%%%%%
\begin{figure}
\ifpdf
\includegraphics[width=\textwidth]{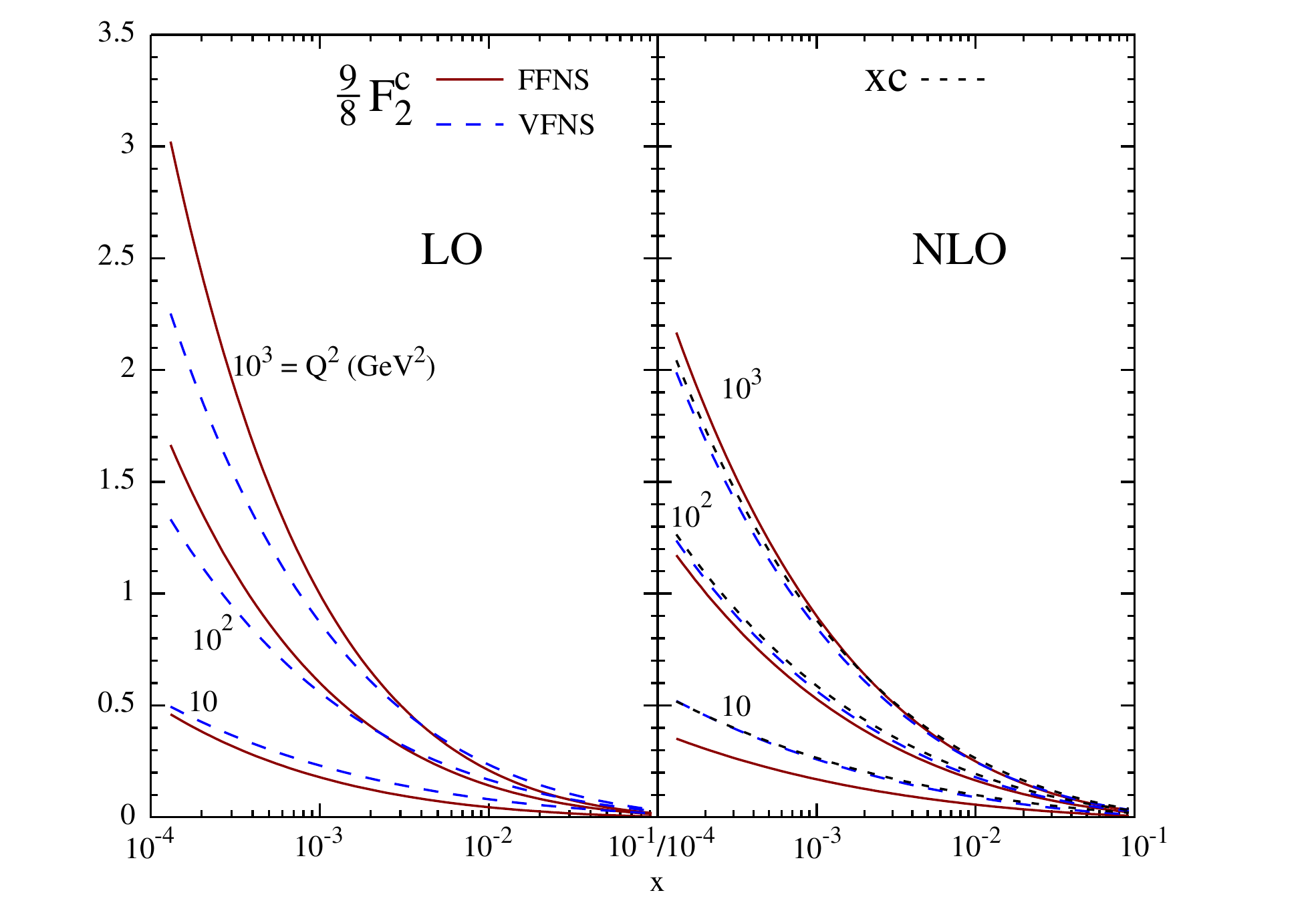}
\else
\includegraphics[width=\textwidth]{DOTH_0801_Fig1.eps}
\fi
%\begin{figure}
%\begin{center}
%\epsfig{figure= DOTH_0801_Fig1.eps}
%\end{center}
\caption{The predicted $x$--dependencies of $\frac{9}{8}F_2^c(x,Q^2)$ in 
the FFNS and VFNS at some typical fixed values of $Q^2$. For the
FFNS the renormalization and factorization scales are chosen to be
$\mu_R^2=\mu_F^2\equiv\mu^2=Q^2+4m_c^2$ with $m_c=1.3$ GeV, and,
as usual, $\mu^2=Q^2$ for the VFNS\@.  The NLO--VFNS charm distribution
is given by $xc(x,Q^2)$ as shown by the short--dashed curves.}
\end{figure}
\clearpage
%%%%%%%%%%%%%%%%%%%%%%%%%%%%%%%%%%%%%%%%%%%%%%%%%%%%%%%%%%%%%%%%%%%%%%%%%%
\begin{figure}
\ifpdf
\includegraphics[width=\textwidth]{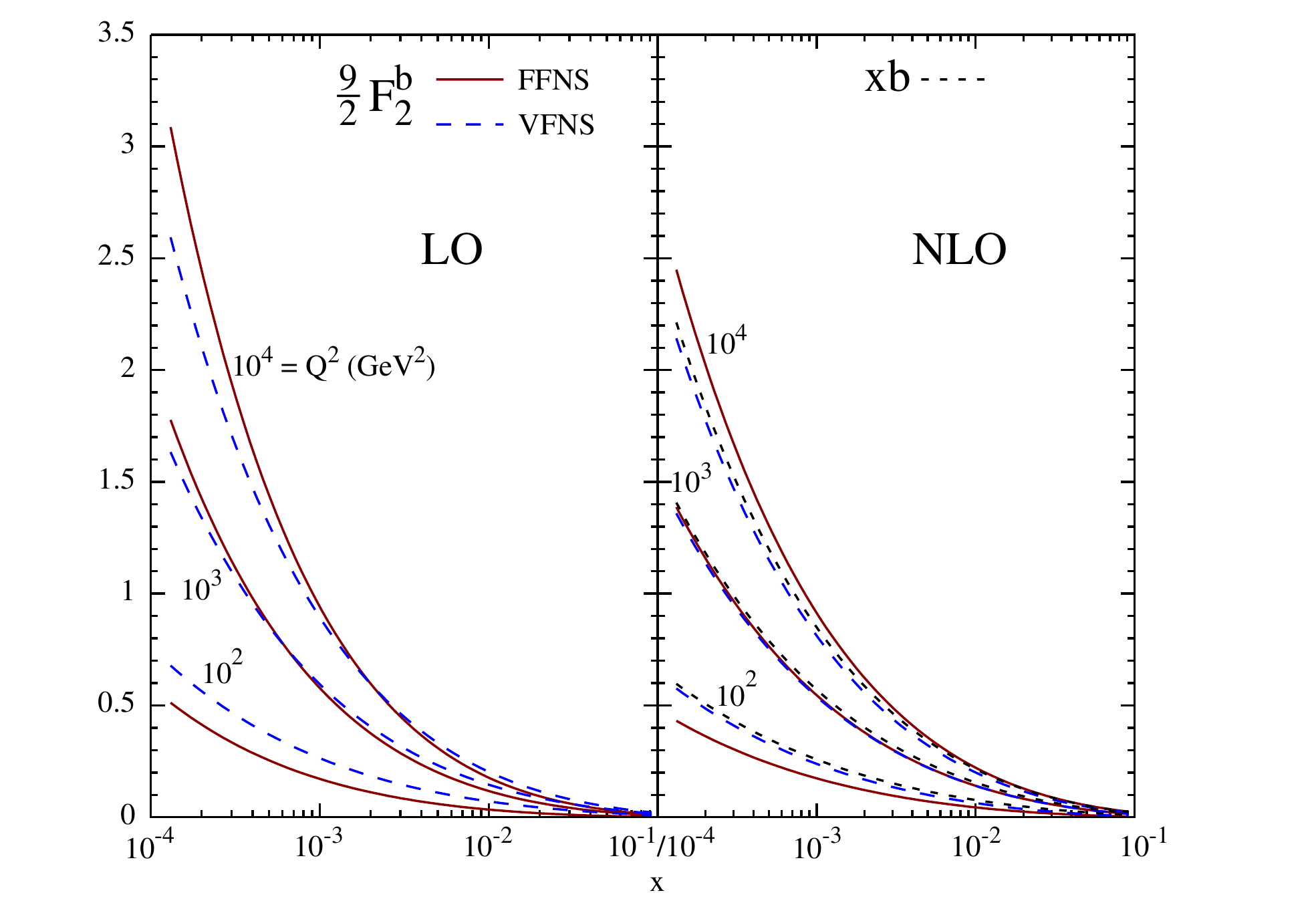}
\else
\includegraphics[width=\textwidth]{DOTH_0801_Fig2.eps}
\fi
%\begin{figure}
%\begin{center}
%\epsfig{figure= DOTH_0801_Fig2.eps}
%\end{center}
\caption{As in Fig.~1 but for bottom production, i.e.\ 
$\frac{9}{2}F_2^b(x,Q^2)$, choosing 
$\mu_R^2 = \mu_F^2\equiv\mu^2=Q^2+4m_b^2$ with $m_b = 4.2$ GeV for the FFNS\@.
The short--dashed curves show the NLO--VFNS bottom distribution 
$xb(x,Q^2)$.}
\end{figure}
\clearpage
%%%%%%%%%%%%%%%%%%%%%%%%%%%%%%%%%%%%%%%%%%%%%%%%%%%%%%%%%%%%%%%%%%%%%%%%%%%%%%
\begin{figure}
\begin{center}
\ifpdf
\includegraphics[width=10.0cm]{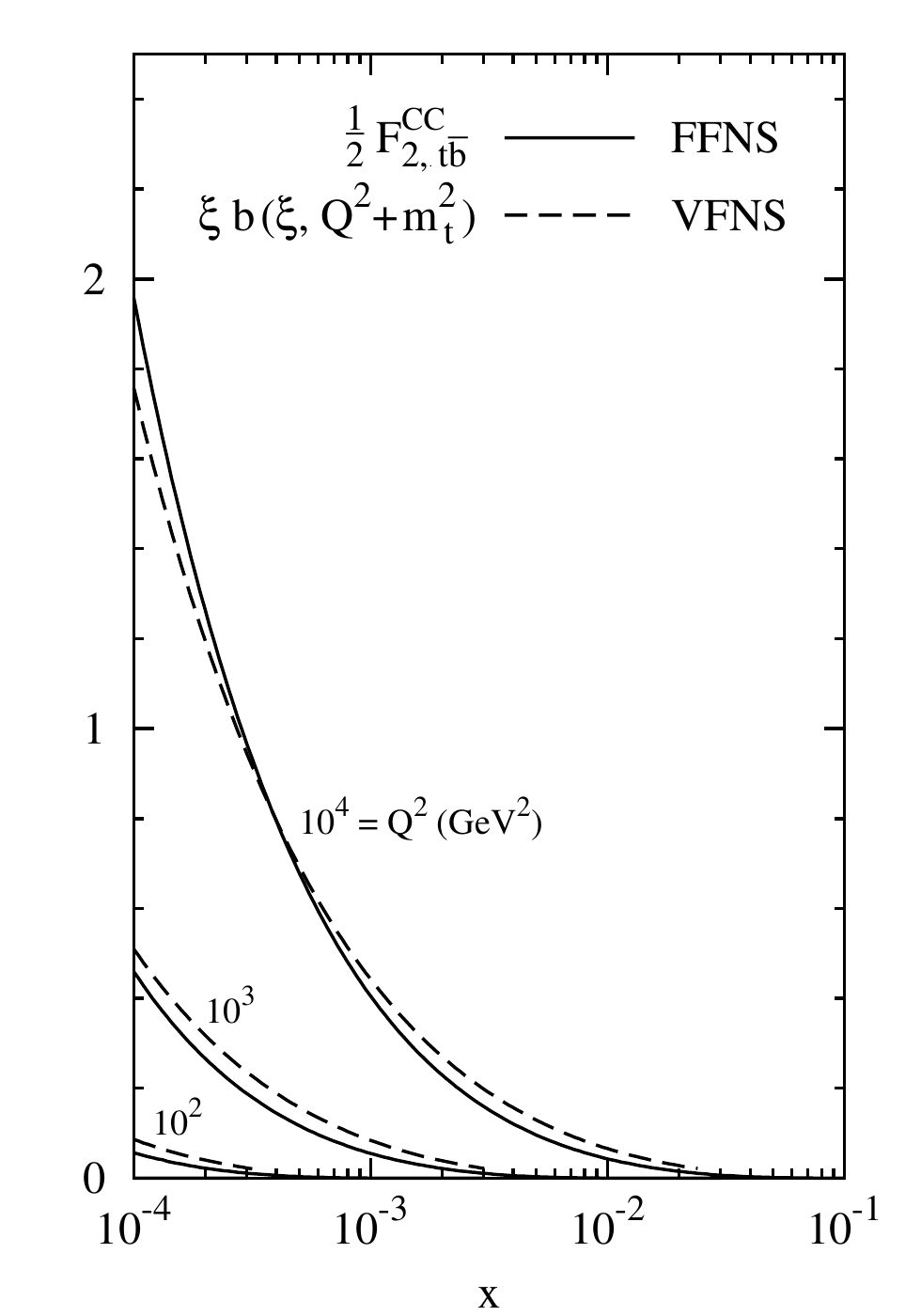}
\else
\includegraphics[width=10.0cm]{DOTH_0801_Fig3.ps}
\fi
\end{center}
%\begin{figure}
%\begin{center}
%\epsfig{figure= DOTH_0801_Fig3.ps}
%\end{center}
\caption{LO predictions for the $x$--dependencies of the weak charged current
structure function $\frac{1}{2}F_{2,t\bar{b}}^{CC}(x,Q^2)$ for $t\bar{b}$
production in the FFNS at some typical fixed values of $Q^2$.  The momentum
scale is chosen to be $\mu_R^2=\mu_F^2\equiv\mu^2=Q^2+(m_t+m_b)^2$ with
$m_t=175$ GeV\@.  These predictions are compared with the bottom distribution
$\xi b(\xi,\, Q^2+m_t^2)$ in the VFNS where $\xi = x(1+m_t^2/Q^2)$.}
\end{figure}
\clearpage
%%%%%%%%%%%%%%%%%%%%%%%%%%%%%%%%%%%%%%%%%%%%%%%%%%%%%%%%%%%%%%%%%%%%%%%%%%%%%%
\begin{figure}
\ifpdf
\includegraphics[width=15.5cm]{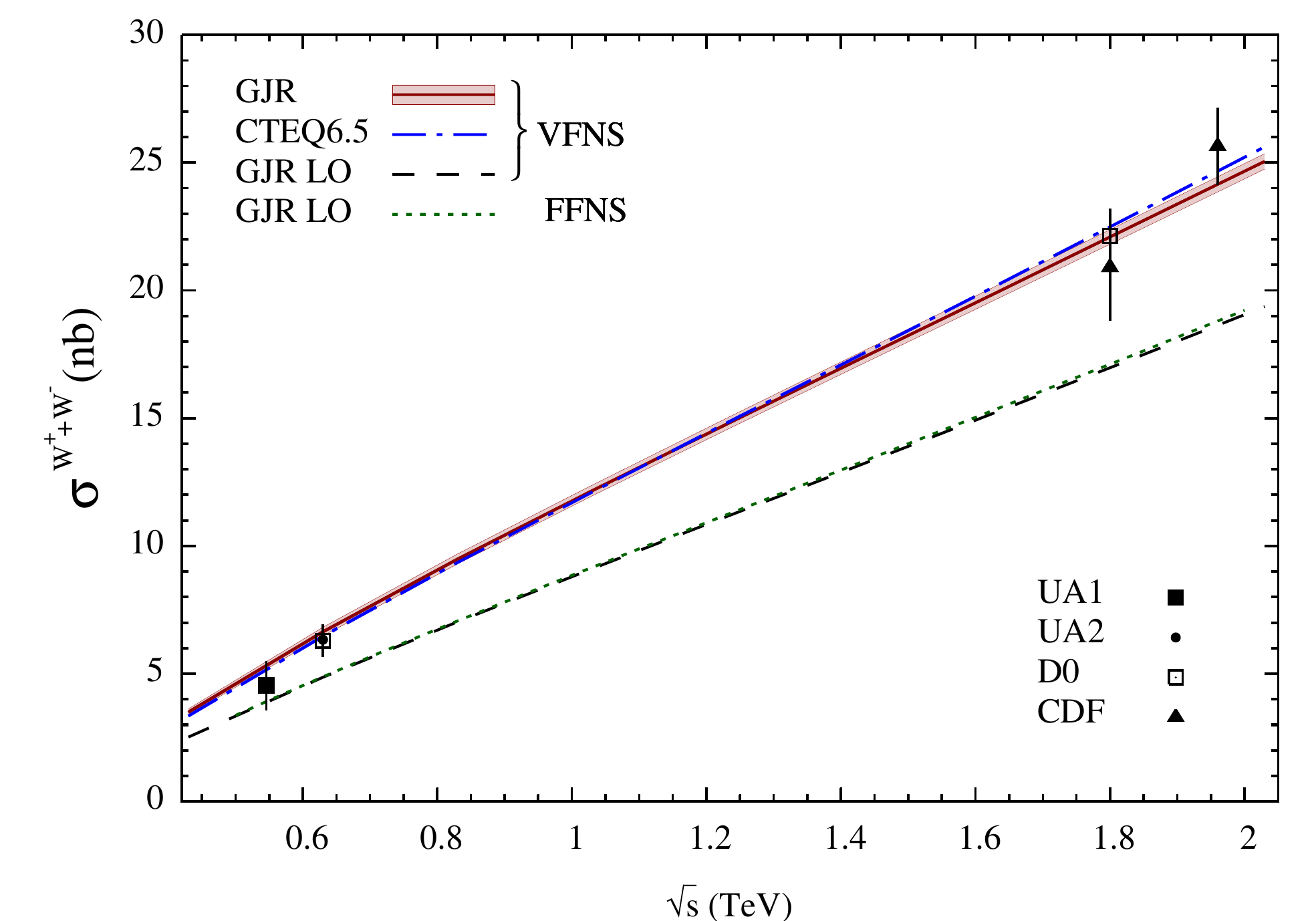}
\else
\includegraphics[width=15.5cm]{DOTH_0801_Fig4.eps}
\fi
%\begin{figure}
%\begin{center}
%\epsfig{figure= DOTH_0801_Fig4.eps, width=13.5cm}
%\end{center}
\caption{Predictions for the total $W^++W^-$ production rates at $p\bar{p}$
colliders with the data taken from \cite{ref10,ref11,ref12,ref13,ref14}.
The LO and NLO GJR parton distributions in the VFNS have been generated from
the FFNS ones \cite{ref1} as described in the text.  The NLO--VFNS
CTEQ6.5 distributions are taken from \cite{ref15}.  The adopted momentum
scale is $\mu_R=\mu_F\equiv\mu=M_W$.  The scale uncertainty of our NLO GJR
predictions, due to varying $\mu$ according to $\frac{1}{2}M_W\leq\mu\leq
2M_W$, amounts to less than 2\% at $\sqrt{s}=1.96$ TeV, for example.  The
shaded region around our central GJR predictions is due to the $\pm 1\sigma$
uncertainty implied by our dynamical NLO parton distributions \cite{ref1}.}
\end{figure}
\clearpage
%%%%%%%%%%%%%%%%%%%%%%%%%%%%%%%%%%%%%%%%%%%%%%%%%%%%%%%%%%%%%%%%%%%%%%%%%%%%
\begin{figure}
\ifpdf
\includegraphics[width=\textwidth]{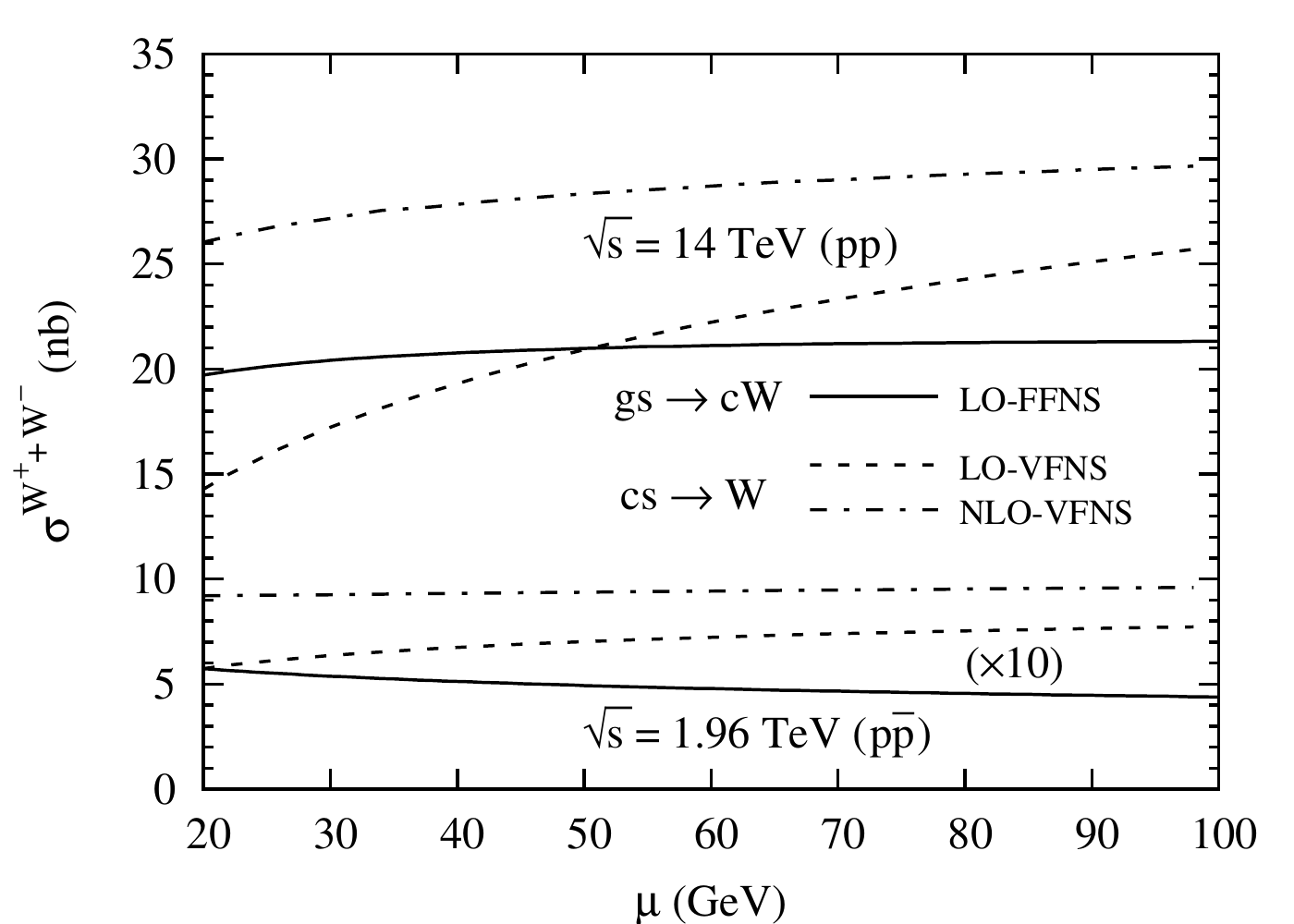}
\else
\includegraphics[width=\textwidth]{DOTH_0801_Fig5.ps}
\fi
%\begin{figure}
%\begin{center}
%\epsfig{figure= DOTH_0801_Fig5.ps}
%\end{center}
\vspace{1.0cm}
\caption{The scale dependence ($\mu_R=\mu_F\equiv\mu$) of the LO--FFNS
contribution to the total $W^+ +W^-$ production rate due to the subprocess
$gs\to cW$ compared to the LO and NLO ones in the VFNS due to $cs\to W$
fusion.  The results refer to the $pp$--LHC ($\sqrt{s}=14$ TeV) and to 
the $p\bar{p}$--Tevatron ($\sqrt{s}=1.96$ TeV) with the latter ones being
multiplied by a factor of 10 as indicated.}
\end{figure}
%%%%%%%%%%%%%%%%%%%%%%%%%%%%%%%%%%%%%%%%%%%%%%%%%%%%%%%%%%%%%%%%%%%%%%%%%%
\begin{figure}
\ifpdf
\includegraphics[width=\textwidth]{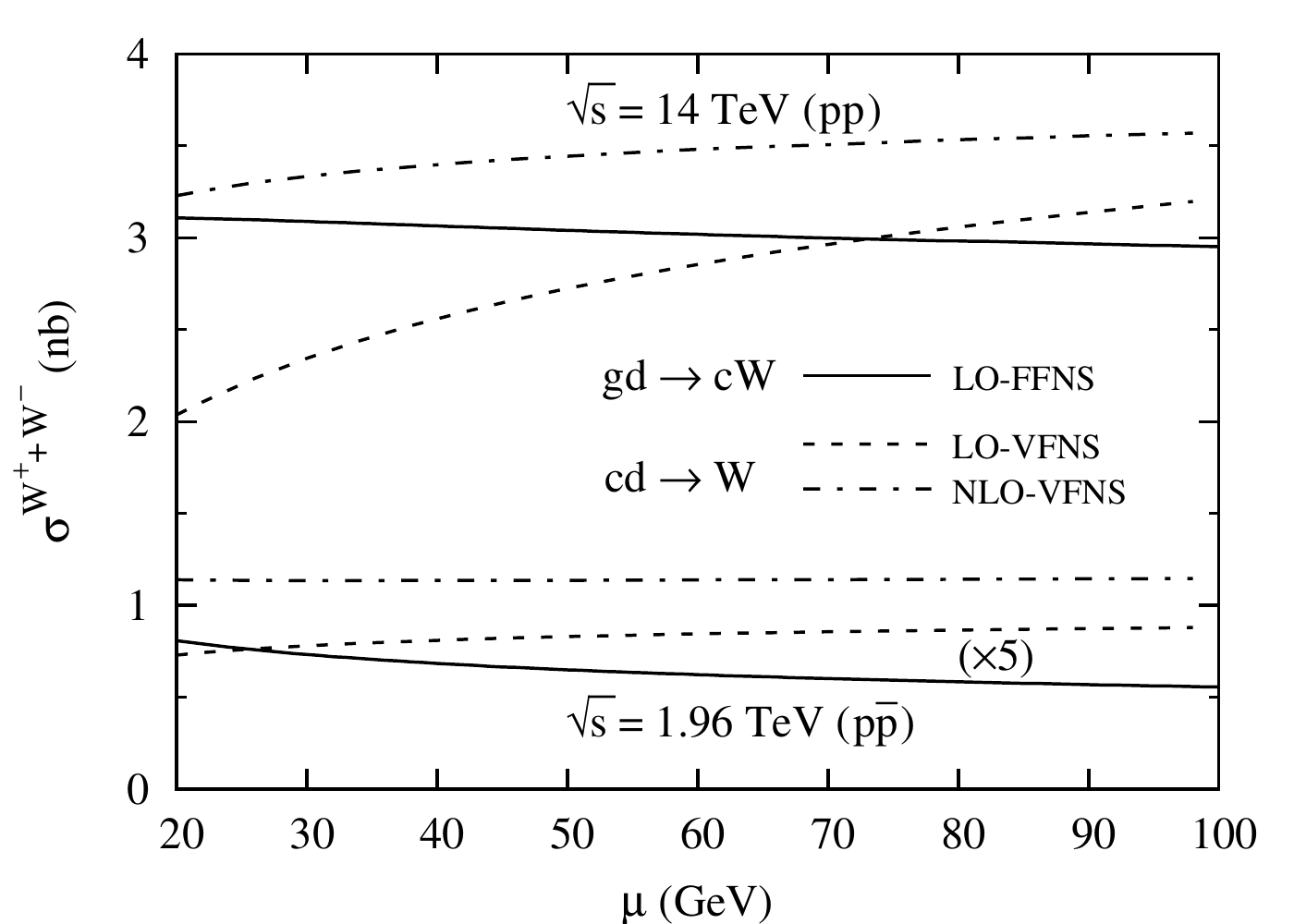}
\else
\includegraphics[width=\textwidth]{DOTH_0801_Fig6.ps}
\fi
%\begin{figure}
%\begin{center}
%\epsfig{figure= DOTH_0801_Fig6.ps}
%\end{center}
\vspace{1.0cm}
\caption{As in Fig.~5 but for the FFNS subprocess $gd\to cW$ to be
compared with $cd \to W$ in the VFNS\@.  The results for the Tevatron 
($\sqrt{s}=1.96$ TeV) are multiplied by a factor of 5 as indicated.}
\end{figure}
%%%%%%%%%%%%%%%%%%%%%%%%%%%%%%%%%%%%%%%%%%%%%%%%%%%%%%%%%%%%%%%%%%%%%%%%%%
%%%%%%resubmission%%%%%%%%%%%%%%%%%%%%%%%%%%%%%%%%%%%%%%%%%%%%%%%%%%%%%%%%%%%%%%%%%%%%
\begin{figure}
\ifpdf
\includegraphics[width=\textwidth]{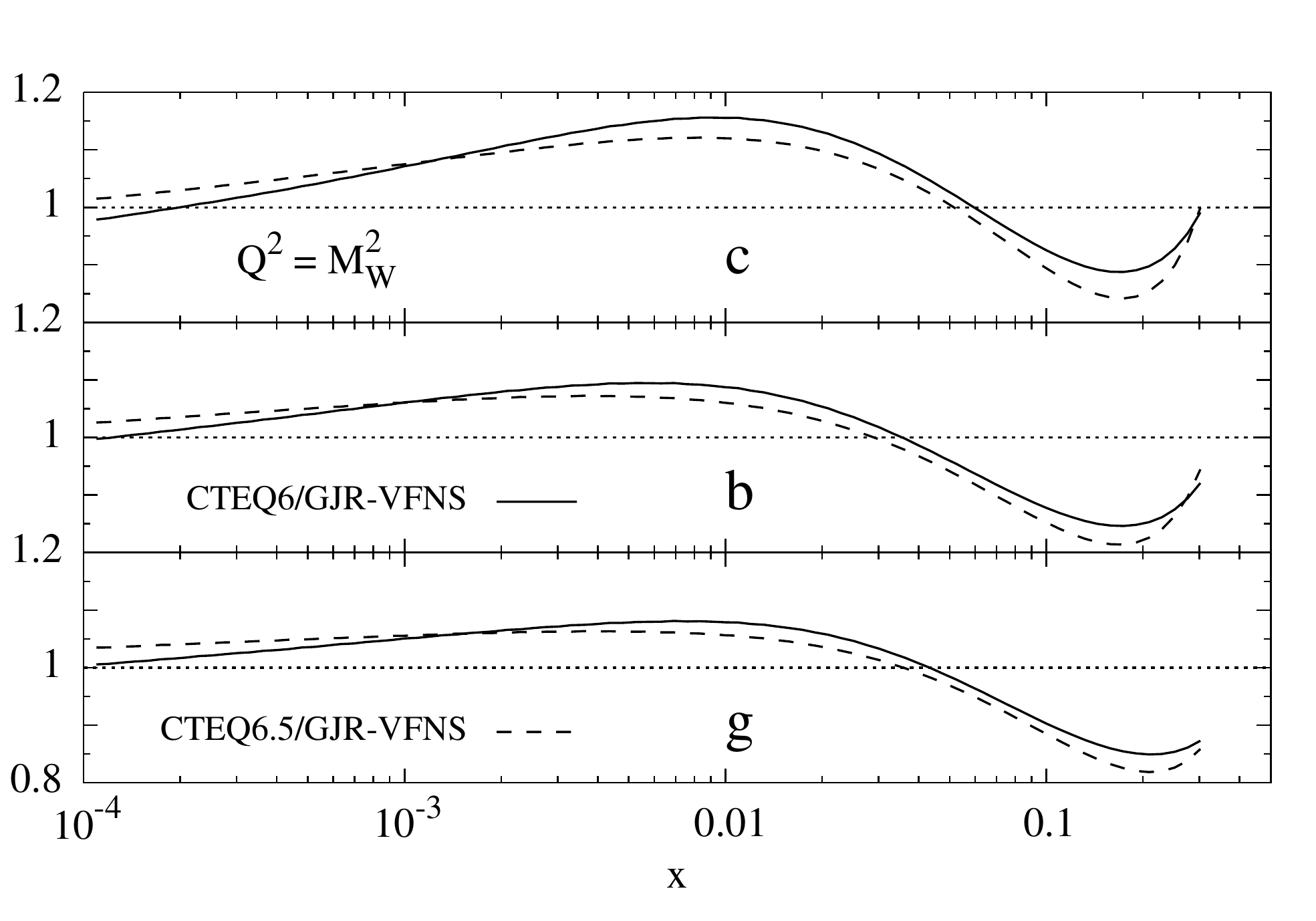}
\else
\includegraphics[width=\textwidth]{DOTH_0801_Fig7.ps}
\fi
%\begin{figure}
%\begin{center}
%\epsfig{figure= DOTH_0801_Fig6.ps}
%\end{center}
\vspace{1.0cm}
\caption{Comparing our present (GJR-VFNS) dynamical parton distributions
generated in the VFNS at NLO($\overline{\rm MS}$) with the ones of
CTEQ6 \cite{ref26} and CTEQ6.5 \cite{ref15} at $Q^2=M_W^2$.}
\end{figure}
%%%%%%%%%%%%%%%%%%%%%%%%%%%%%%%%%%%%%%%%%%%%%%%%%%%%%%%%%%%%%%%%%%
\end{document}